\documentclass[aps,pre,twocolumn,groupedaddress,footinbib]{revtex4-1}

\usepackage{color}
\usepackage{amsmath}
\usepackage[pdftex]{graphicx}
\usepackage{subfigure}

\begin{document}

\title{Aging-induced continuous phase transition}

\author{Oriol Artime}
\email[]{oriol@ifisc.uib-csic.es}
\author{Antonio F. Peralta}
\author{Ra\'ul Toral}
\author{Jos\'e J. Ramasco}
\author{Maxi San Miguel}
\affiliation{Instituto de F\'isica Interdisciplinar y Sistemas Complejos IFISC (CSIC-UIB), Campus UIB, 07122 Palma de Mallorca, Spain}

\begin{abstract}
Aging is considered as the property of the elements of a system to be less prone to change states as they get older. We incorporate aging into the noisy voter model, a stochastic model in which the agents modify their binary state by means of noise and pair-wise interactions. Interestingly, due to aging the system passes from a finite-size discontinuous transition between ordered (ferromagnetic) and disordered (paramagnetic) phases to a second order phase transition, well defined in the thermodynamic limit, belonging to the Ising universality class. We characterize it analytically by finding the stationary solution of an infinite set of mean field equations. The theoretical predictions are tested with extensive numerical simulations in low dimensional lattices and complex networks. We finally employ the aging properties to understand the symmetries broken in the phase transition.
\end{abstract}

\maketitle

\section{Introduction}
Stochastic binary-state models are a versatile tool to describe a large variety of natural phenomena. The individual elements of a system are given a state that evolves via interactions with their neighbors. This framework is rather 
general, and it has been used to model many systems, such as magnetic materials \cite{baxter2016exactly}, percolation \cite{stauffer1994introduction}, epidemic spreading \cite{keeling2008modeling,pastor2015epidemic}, neural activity \cite{bogacz2006physics, hopfield1982neural}, language dynamics \cite{abrams2003linguistics, patriarca2012modeling} or economics \cite{kirman1993ants, lux1999scaling}. All these spin-like, agent-based models extract the basic features of the phenomena they want to describe. Extensions or modifications of the dynamical rules sometimes lead to dramatic changes with respect to the original models. For example, the inclusion of temporal correlations in the activation of the elements of a system \cite{artime2017dynamics,goh2008burstiness,karsai2012universal}, the role of noise \cite{broeck1994noise}, or the presence of nontrivial structures in the connectivity, such as graphs formed by communities \cite{onnela2007structure, masuda2014voter} or multilayer networks \cite{artime2017joint,diakonova2016irreducibility,klimek2016dynamical,domenico2016physics} bring staggering new dynamical effects.

\begin{figure*}
\minipage{0.33\textwidth}
  \includegraphics[width=\linewidth]{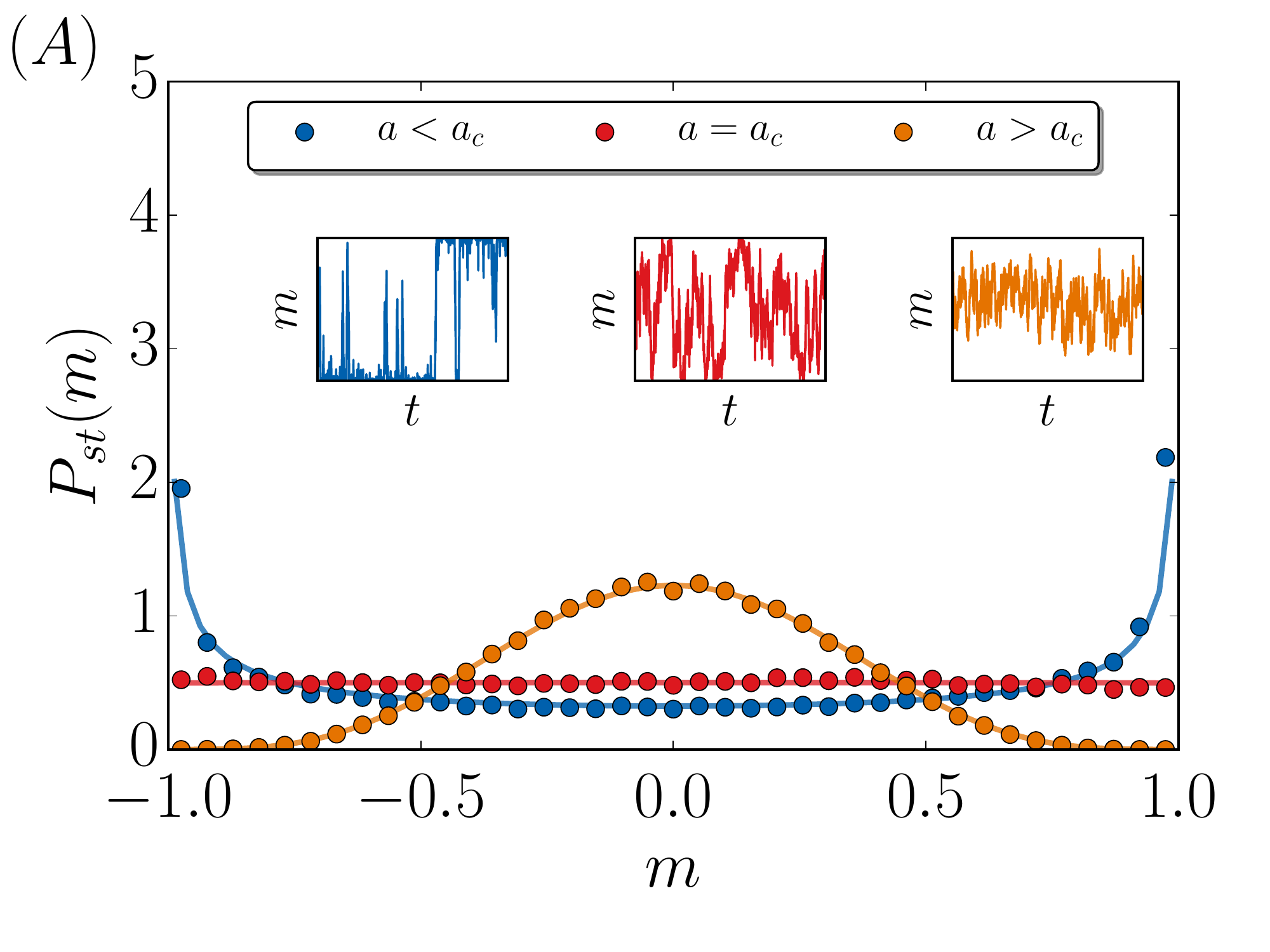}
\endminipage\hfill
\minipage{0.33\textwidth}
  \includegraphics[width=\linewidth]{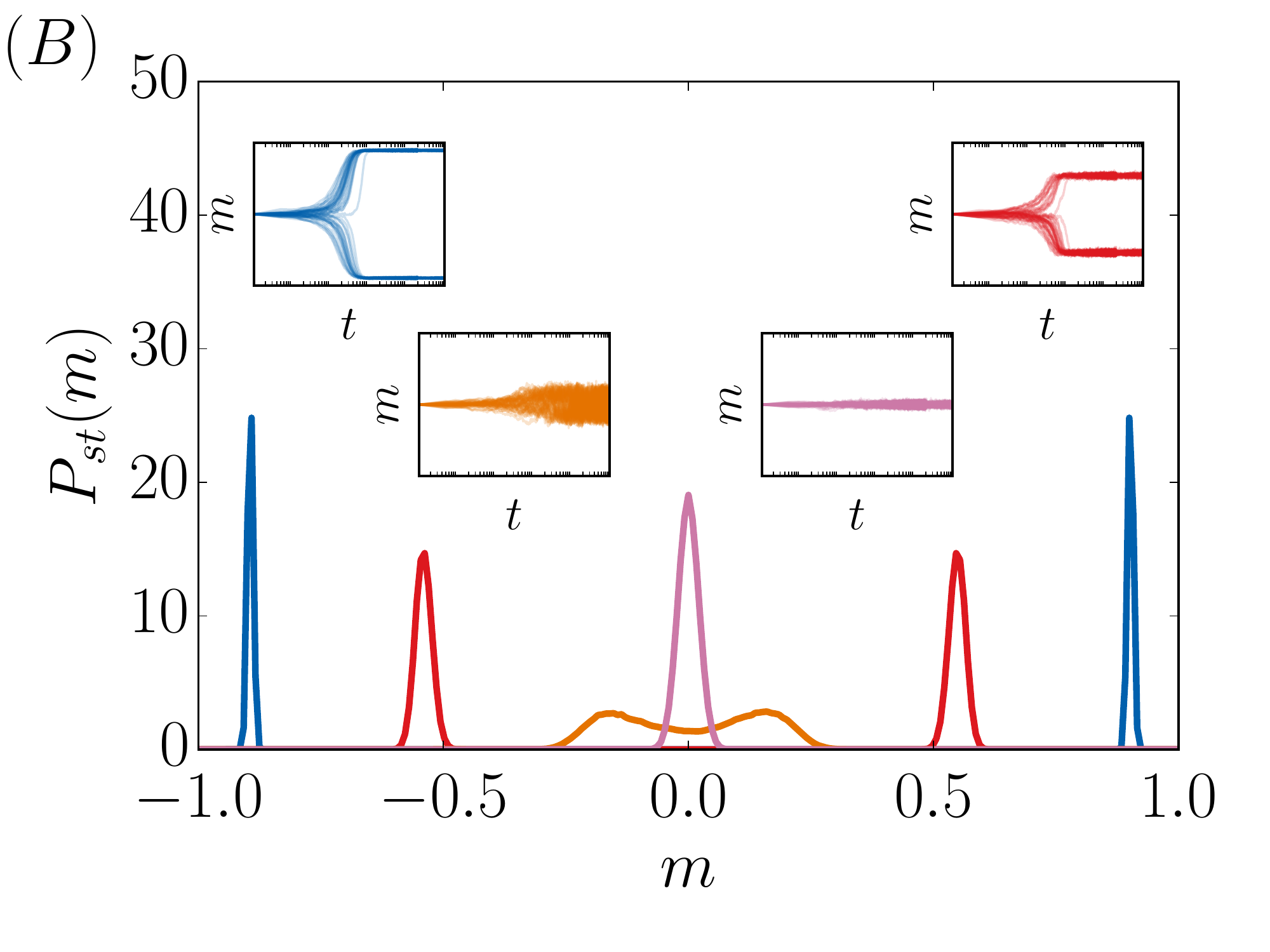}
\endminipage\hfill
\minipage{0.33\textwidth}%
  \includegraphics[width=\linewidth]{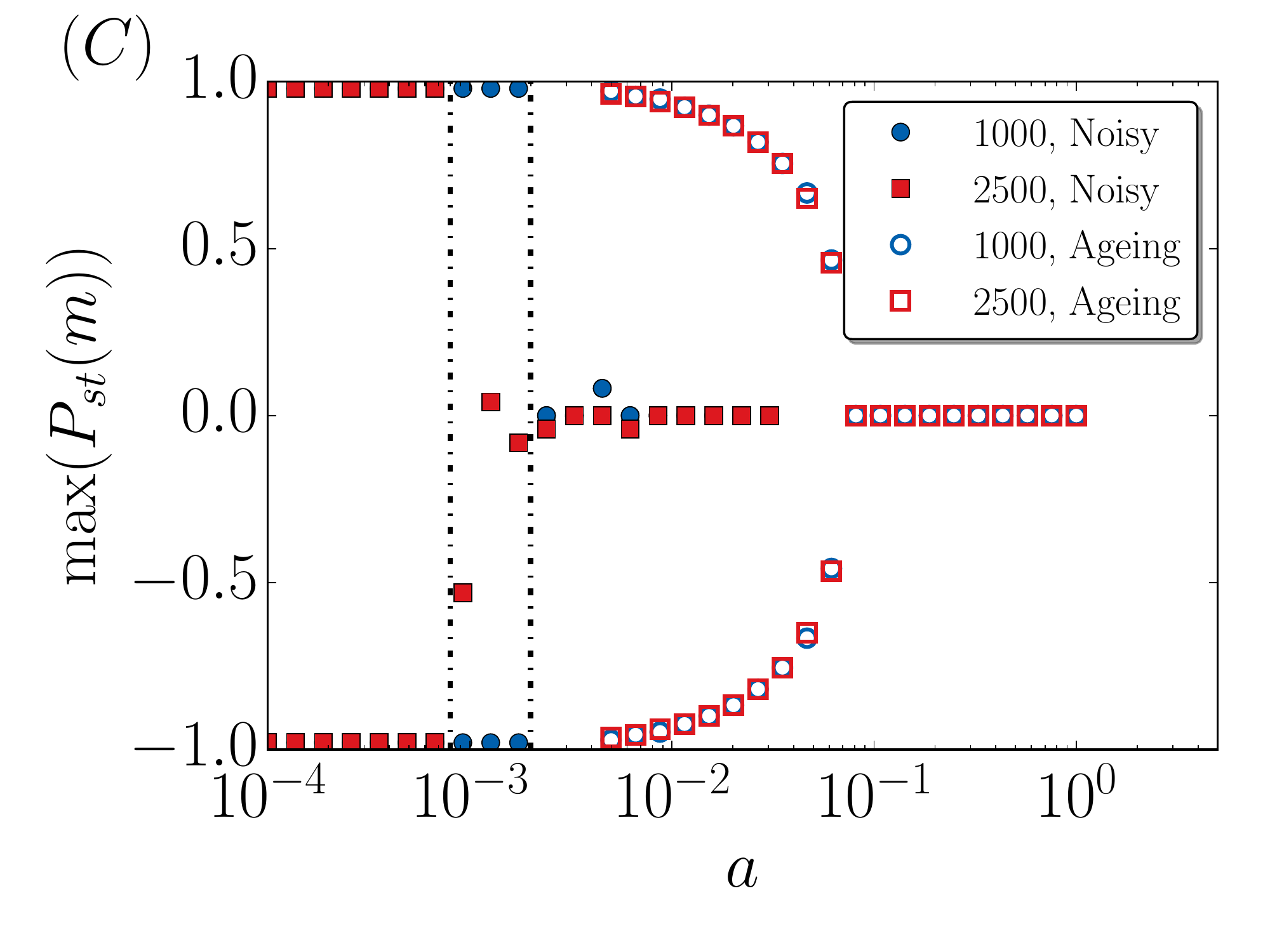}
\endminipage
\caption{Numerical simulations of ($A$) the noisy voter model and ($B$) the noisy voter model with aging. In $(A)$, the stationary probability density function (pdf) of the magnetization in the three different regimes. The points come from simulations, and the solid lines are the theoretical curves, Eq.~\eqref{eq:app1_4}. The insets show one typical trajectory of the dynamics in each of the regimes. In $(B)$, the stationary pdf for the noisy voter model with aging in the different regimes. The insets show $50$ trajectories of the magnetization. The considered range of the insets is always $ [-1,1] $. In $(C)$, the maxima of the pdf of the two models as a function of the noise parameter $ a $ for different system sizes. The filled symbols correspond to $(A)$, the empty symbols are for $(B)$. In the latter case, the points corresponding to the two sizes overlap so much that they are almost indistinguishable from one another. The vertical lines are the predictions for the critical noise of the noisy voter model without aging for the two sizes considered in the plot.}\label{fig:Fig1}
\end{figure*}

The implications of aging have been widely explored in different areas of research, although with different meanings. In computational biology literature, aging is taken as the increase in the mortality of a species as its population gets older \cite{azbel1996unitary,azbel1997phenomenological,penna1995bit}. In nonequilibrium statistical physics, aging appears when the relaxation to the stationary state of a system displays slow dynamics (i.e., it is nonexponential), there is dynamical scaling, and the time-translational invariance is broken \cite{henkel2011non}. In chemistry, chemical aging comes to the fore when the properties of a material change over time without any forces acting on it but due to slow reactions with the surroundings \cite{henkel2011non}, such as thermal degradation \cite{clough1981combined,cortizo2004effect} or photo-oxidation \cite{robinson2007rethinking,sarna2003loss}. Here we take the approach of considering aging as the influence that persistence times have on the state transitions in a system \cite{fernandez2011update, perez2016competition, boguna2014simulating}. Aging (also called inertia in Ref. \cite{stark2008decelerating}) constrains the transitions in a way that the longer an element remains in a given state, the smaller the probability to change it.

We study the effect of this type of aging in the noisy voter model, also known as the Kirman model \cite{kirman1993ants}. It has appeared in several contexts, such as percolation \cite{lebowitz1986percolation}, surface-catalytic reactions \cite{fichthorn1989noise, considine1989comment}, probability theory \cite{granovsky1995noisy}, opinion dynamics \cite{carro2016noisy,diakonova2015noise,rozanova2017dynamical,khalil2018zealots, peralta2018analytical}, and economics \cite{kirman1993ants,carro2015markets,alfarano2008time}. 
The dynamics of this binary-state model is driven by two mechanisms: noise, defined as spontaneous state changes at a rate $a$, and a pair-wise interaction by which an element blindly copies the state of a randomly chosen neighbor. With these simple ingredients, the system displays a discontinuous finite-size transition as a function of the control parameter $a$. The transition point depends inversely on the system size and, hence, it is located at $a=0$ in the thermodynamic limit. Statistical and critical properties of the transition have been studied in Ref. \cite{lebowitz1986percolation}. 

In this paper we describe the rich phenomenology introduced by aging, which transmutes the nature of the transition to second order, makes it fall into the Ising universality class and, more importantly, places it at a finite $a$ in the thermodynamic limit. We are able to compute the value of the mean magnetization in the stationary regime in the well-mixed scenario, hence finding an expression for the critical point as well as the magnetization critical exponent. The characterization of the phase transition is completed by numerically studying the system in other embedding dimensions than mean field and obtaining other critical exponents. Finally, we exploit the aging properties to give an alternative characterization of the phase transition with an order parameter based on the age of the elements of the system.

The paper is organized as follows. In the next section we describe the original version of the noisy voter model, and we give a brief overview of its properties. Section~\ref{sec:aging} is the core of the paper: the noisy voter model with aging is introduced, the analytical calculations are carried out, and the numerical results are reported and contrasted with the theoretical predictions. The final section contains a summary and a discussion of the results.

\section{Standard noisy voter model}
\label{sec:standard}

We introduce first the standard noisy voter model, which acts as a base case for comparison with its version with aging. Let $ N $ be the number of elements of a system, the so-called agents or nodes, each of them endowed with a binary variable $ s_{j} = \{0,1\} $. The possible contributions to the change of $ s_j $ are due to either noise, in the form of random flips $ s_j \to - s_j $, or a pair-wise interaction with one of $ j $'s neighbors (voter update) where node $ j $ randomly chooses one of her connections and adopts that state. Let $ n = \sum_j s_j$ be the total number of agents in state $ 1 $. The microscopic transition rates for each agent $ j $ in an all-to-all topology can be written as
\begin{align}
\label{eq:1}
\omega_j^+ & \equiv \omega (s_j = 0 \to s_j = 1) = \frac{a}{2} + (1-a) \frac{n}{N}, \nonumber \\
\omega_j^- & \equiv \omega (s_j = 1 \to s_j = 0) = \frac{a}{2} + (1-a) \frac{N-n}{N}.
\end{align}
The first term on the r.h.s. is the contribution of the noise. At each interaction, with probability $ a $ the agent is chosen to perform a noisy update, resulting in state $1$ half of the times and state $ 0 $ the other half, regardless of the former state. With the complementary probability $ 1 - a $, the voter update is performed. The case of the classical voter model is recovered when $ a = 0 $ \cite{liggett2013stochastic}. By writing the global rates in a master equation and expanding them in the inverse of the system size $ N $, one obtains a Fokker--Planck equation for the probability density function $ P(m,t) $,
\begin{equation}
\label{eq:app1_2}
\frac{\partial P(m,t)}{\partial t} = - \frac{\partial }{\partial m} \left[ A(m) P(m,t)\right] + 
\frac{1}{2} \frac{\partial^2 }{\partial m^2} \left[ B(m) P(m,t) \right],
\end{equation}
with drift $ A(m) = - a \, m $ and diffusion $B(m) = 2 \left( a + (1-a)(1-m^2) \right)/N$, where $ m = 2\,n/N - 1 \in [-1,1] $ is the magnetization. This equation can be solved explicitly \cite{artime2018universality}, but valuable information can be extracted from the stationary solution too. Setting the time derivative to $ 0 $, one gets
\begin{equation}
\label{eq:app1_4}
P_{st} (m) = \mathcal{Z}^{-1} \left[ 1 + (a-1)m^2 \right]^\frac{2-a(N+2)}{2(a-1)},
\end{equation}
with the normalization constant being $\mathcal{Z} = 2 _2F_1 (\frac{1}{2}, 1 + \frac{aN}{2(a-1)}, \frac{3}{2}, 1-a) $, where $ _2F_1 $ is the hypergeometric function. The sign of the exponent of the steady state solution changes the convexity of the function. If it is positive, the solution becomes convex with two peaks at the borders of the interval $ m = \pm 1 $. On the contrary, if it is negative we encounter a concave solution with one maximum in the center of the interval ($ m = 0 $, equal coexistence of states). This transition occurs at $ a_c = 2/(N+2) $, and precisely at this value the stationary probability density function is flat, meaning that any magnetization is equiprobable. These three regimes are shown with numerical simulations in Fig.~\ref{fig:Fig1}A. The maximum (maxima) of $P_{st}(m)$ can be used to appreciate the discontinuous nature of the transition as well as the size--dependence of the critical point position [Fig.~\ref{fig:Fig1}C].

\begin{figure*}
\minipage{0.33\textwidth}
  \includegraphics[width=\linewidth]{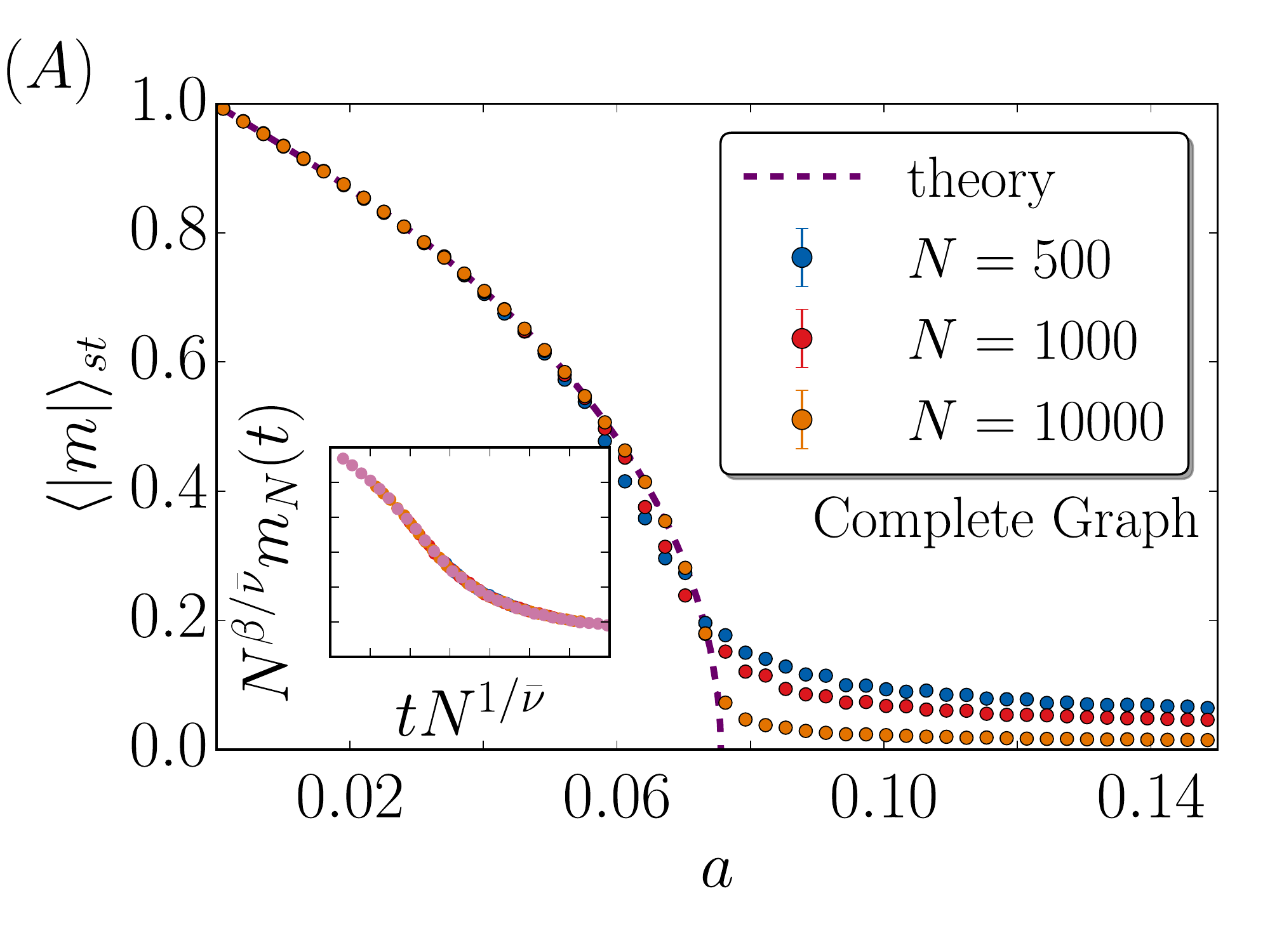}
\endminipage\hfill
\minipage{0.33\textwidth}
  \includegraphics[width=\linewidth]{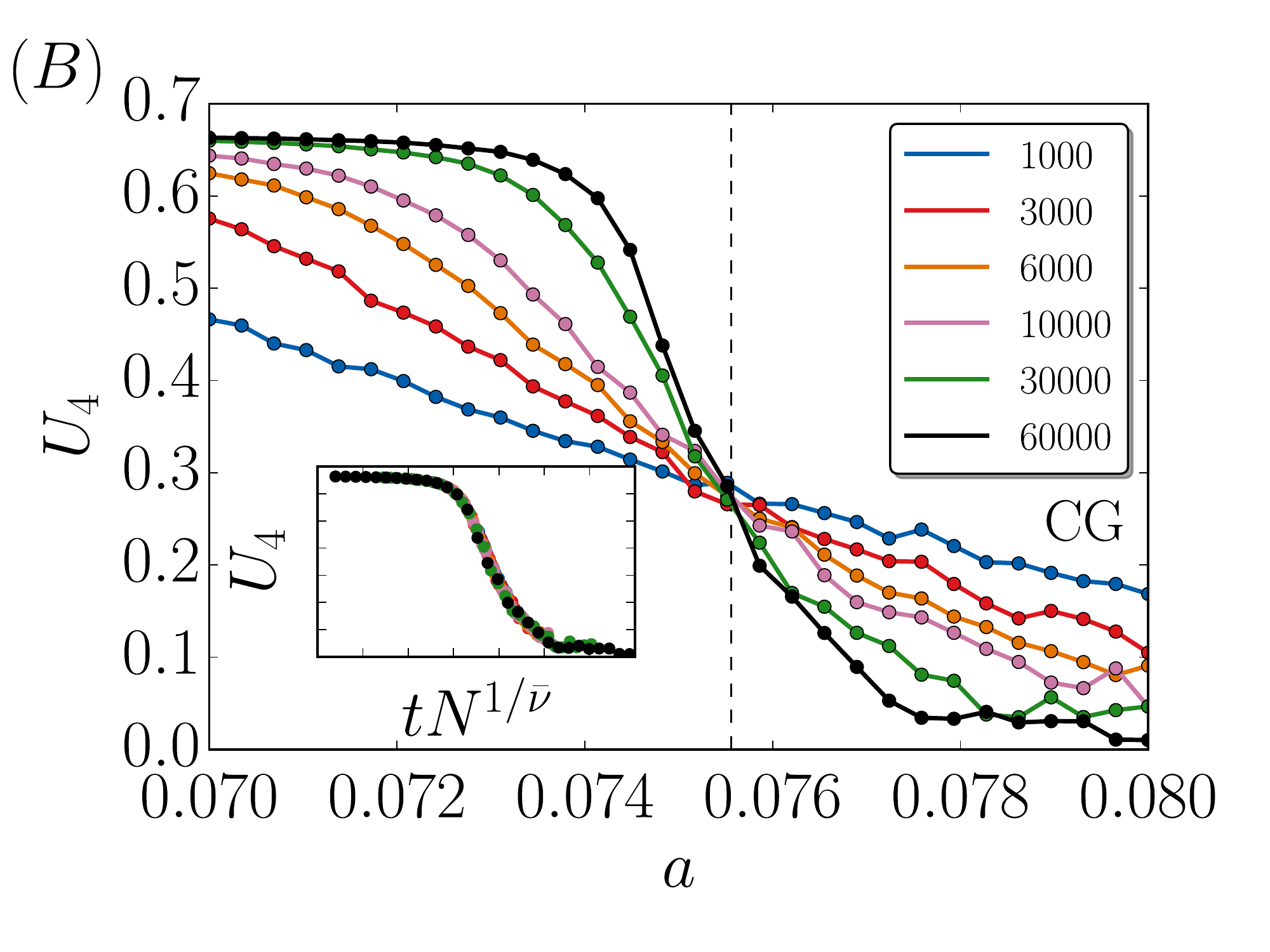}
\endminipage\hfill
\minipage{0.33\textwidth}%
  \includegraphics[width=\linewidth]{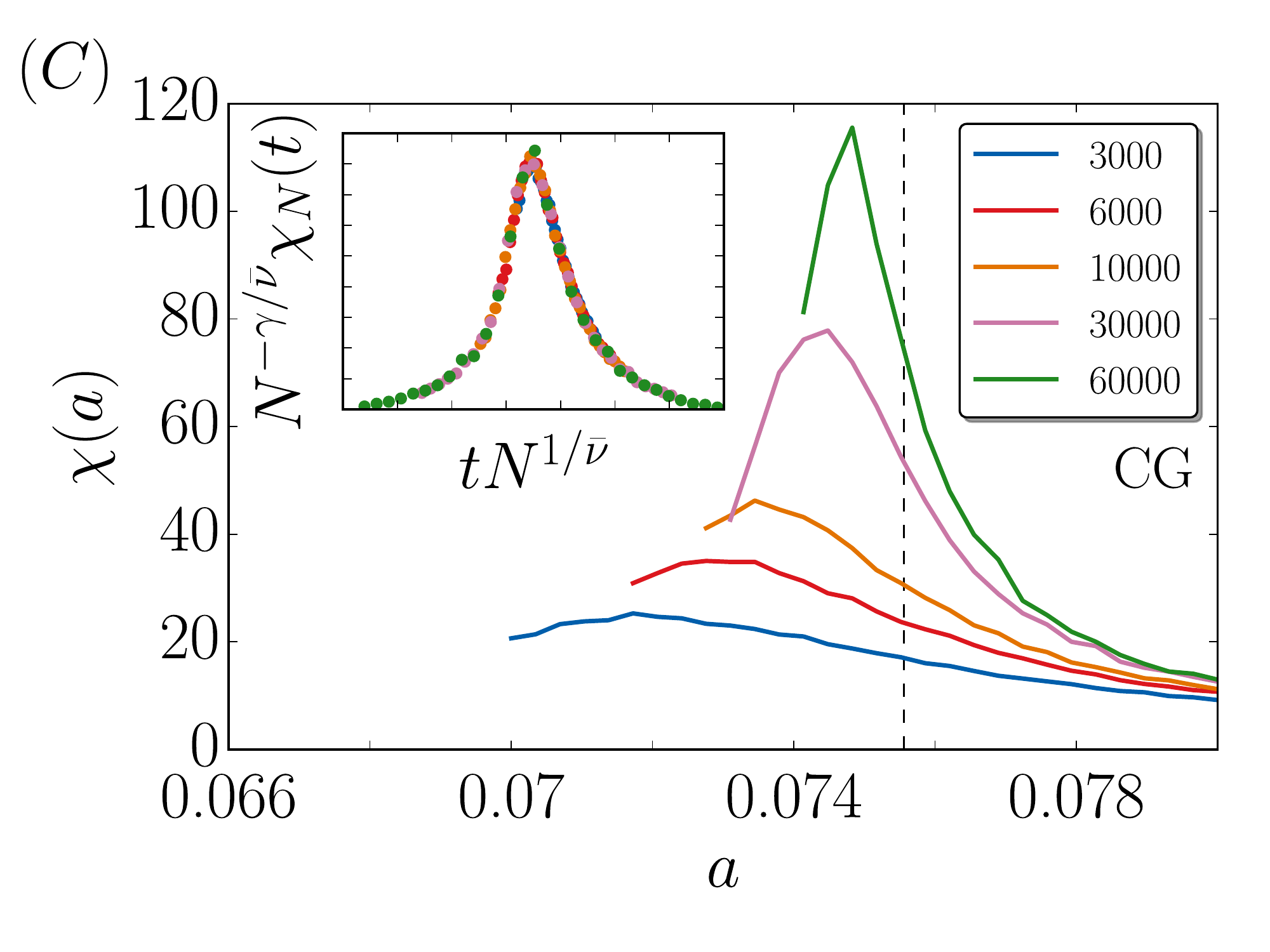}
\endminipage

\caption{($A$) Stationary magnetization [analytical expression Eq.~(\ref{eq:6}) and simulations], ($B$) Binder cumulant, and ($C$) susceptibility for the noisy voter model with aging for different system sizes in a complete graph (CG). The insets show the collapses with the corresponding Ising critical exponents \cite{yeomans1992statistical}. The vertical lines are located at the critical point to guide the eye.
}\label{fig:Fig2}  
\end{figure*}

\begin{figure*}

\minipage{0.33\textwidth}
  \includegraphics[width=\linewidth]{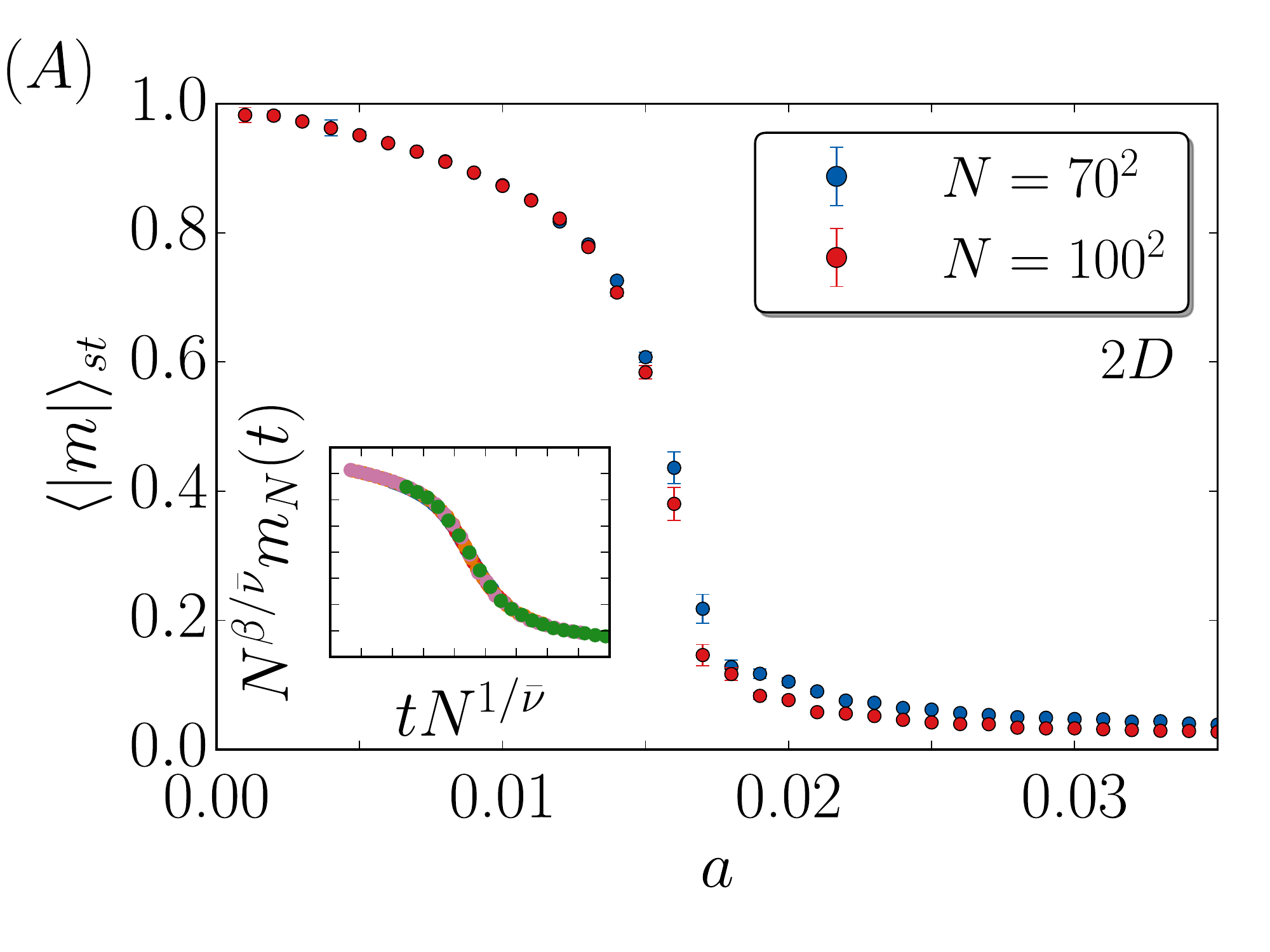}
\endminipage\hfill
\minipage{0.33\textwidth}
  \includegraphics[width=\linewidth]{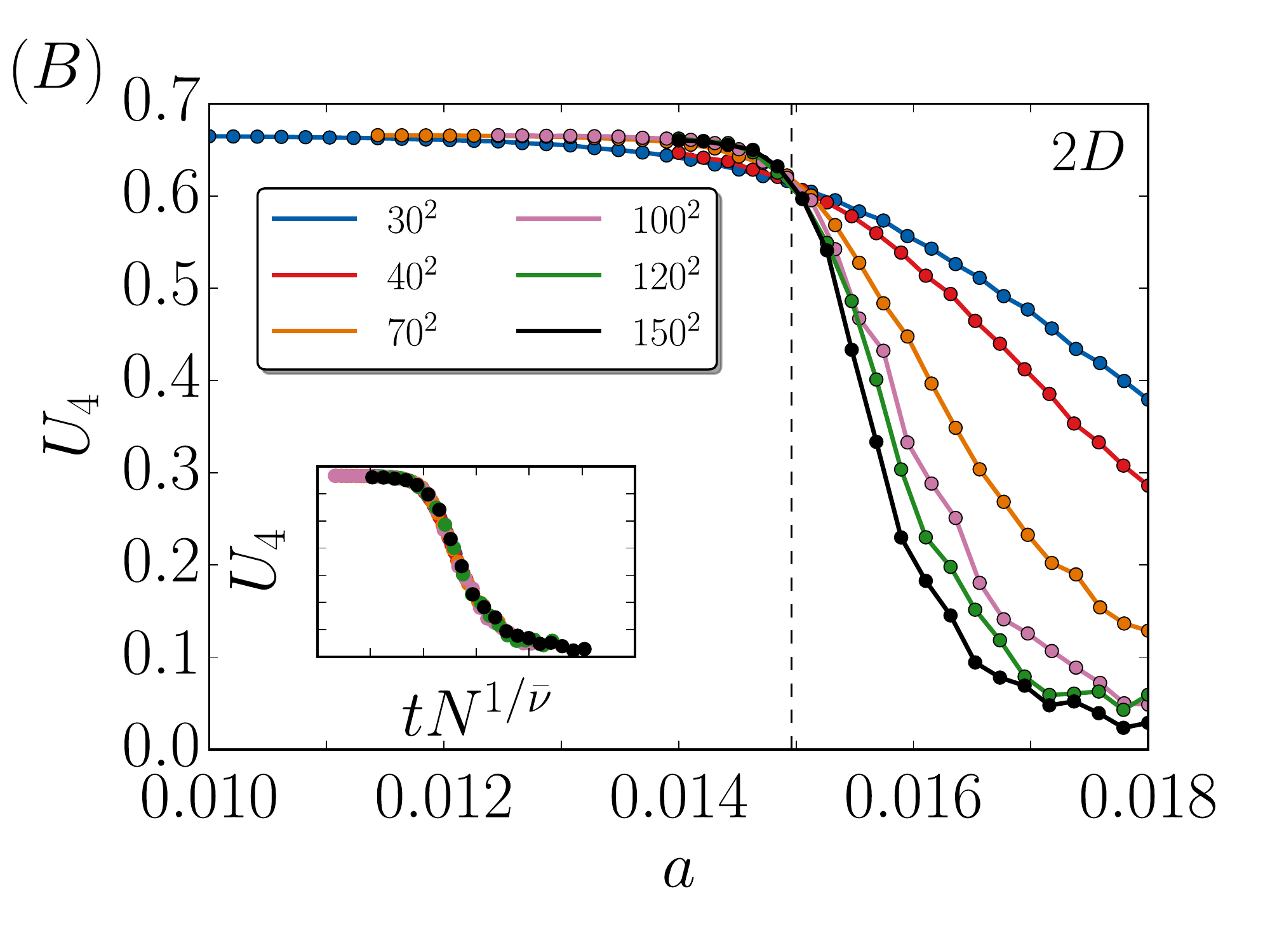}
\endminipage\hfill
\minipage{0.33\textwidth}%
  \includegraphics[width=\linewidth]{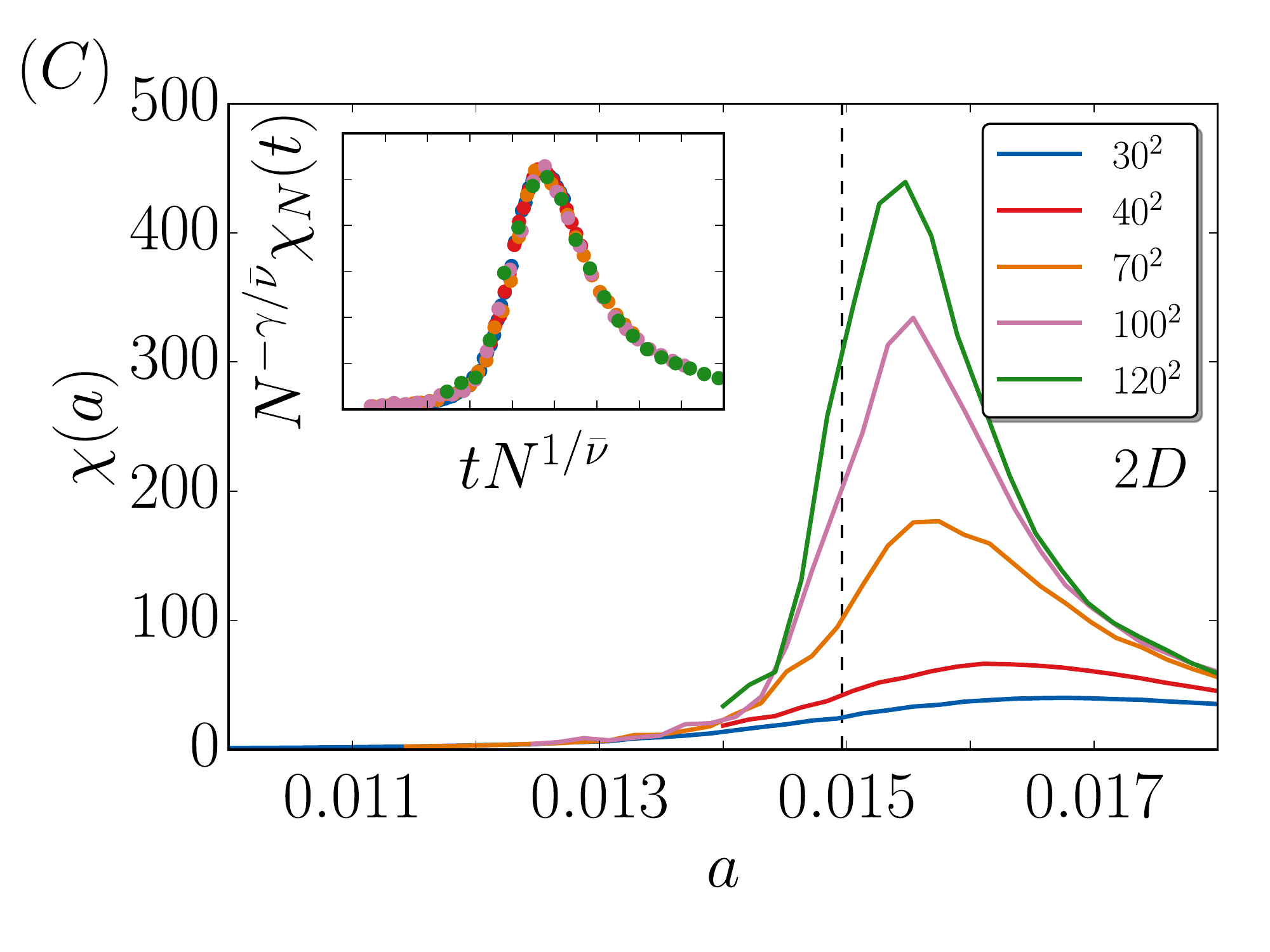}
\endminipage

\minipage{0.33\textwidth}
  \includegraphics[width=\linewidth]{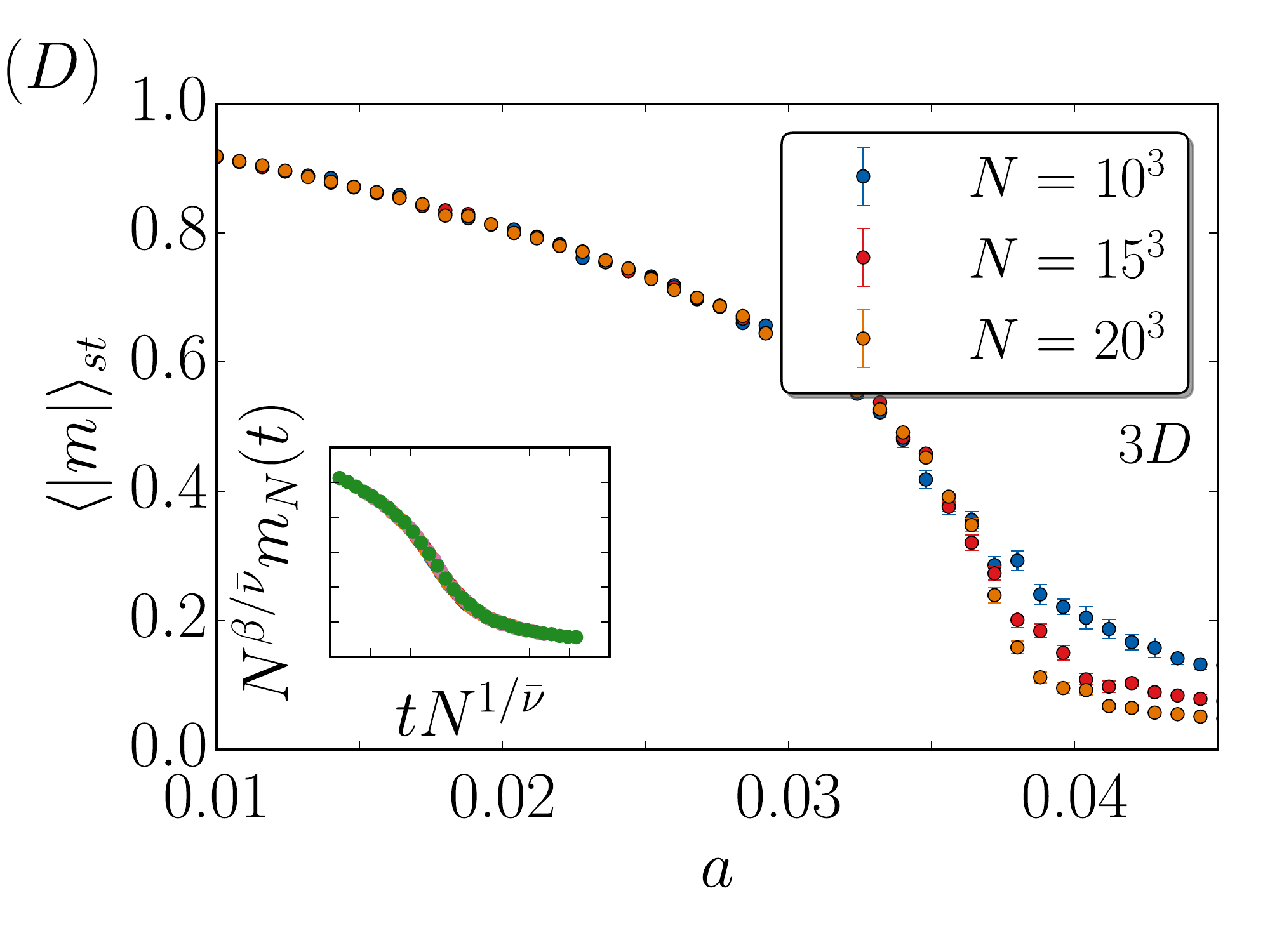}
\endminipage\hfill
\minipage{0.33\textwidth}
  \includegraphics[width=\linewidth]{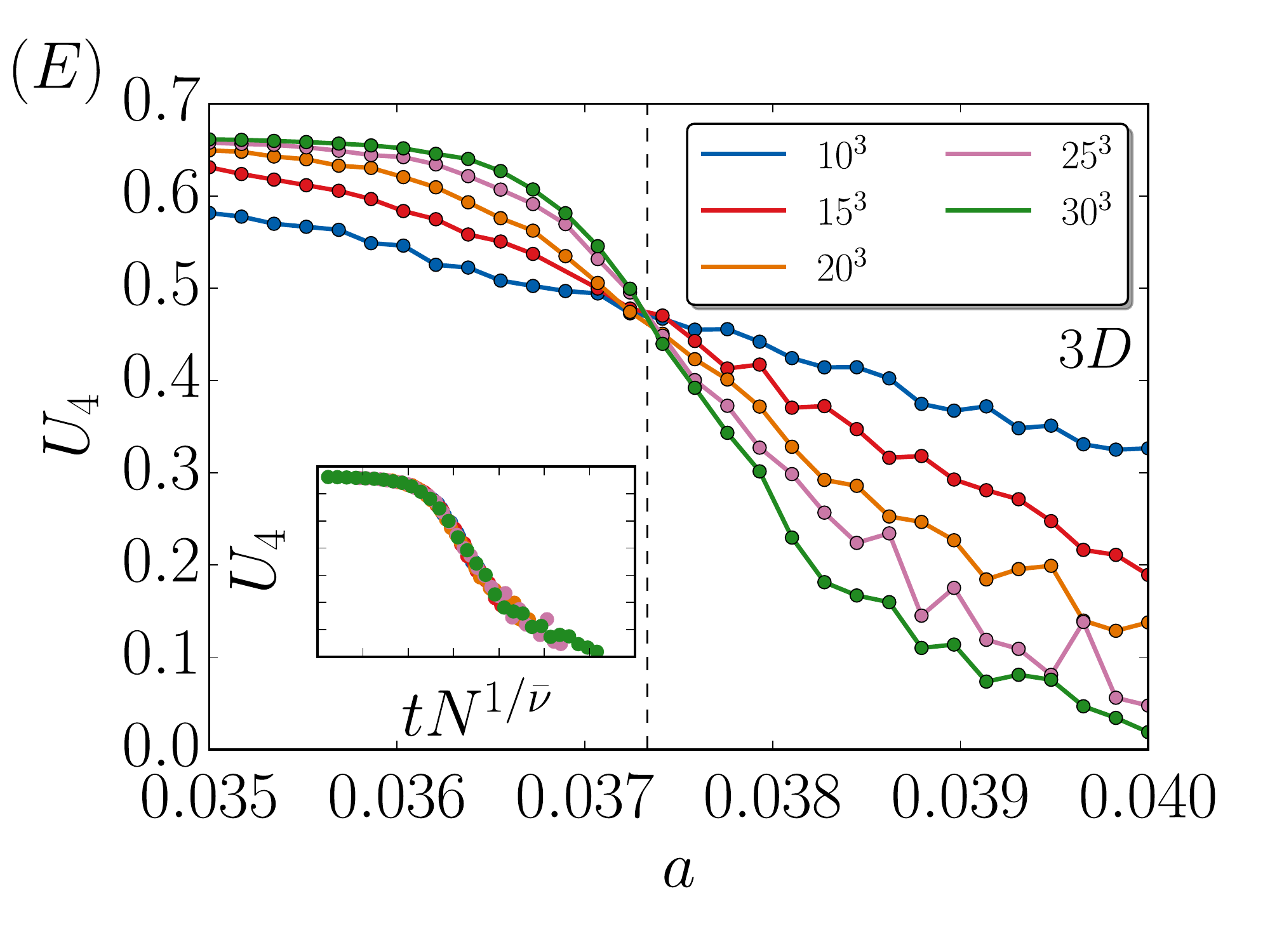}
\endminipage\hfill
\minipage{0.33\textwidth}%
  \includegraphics[width=\linewidth]{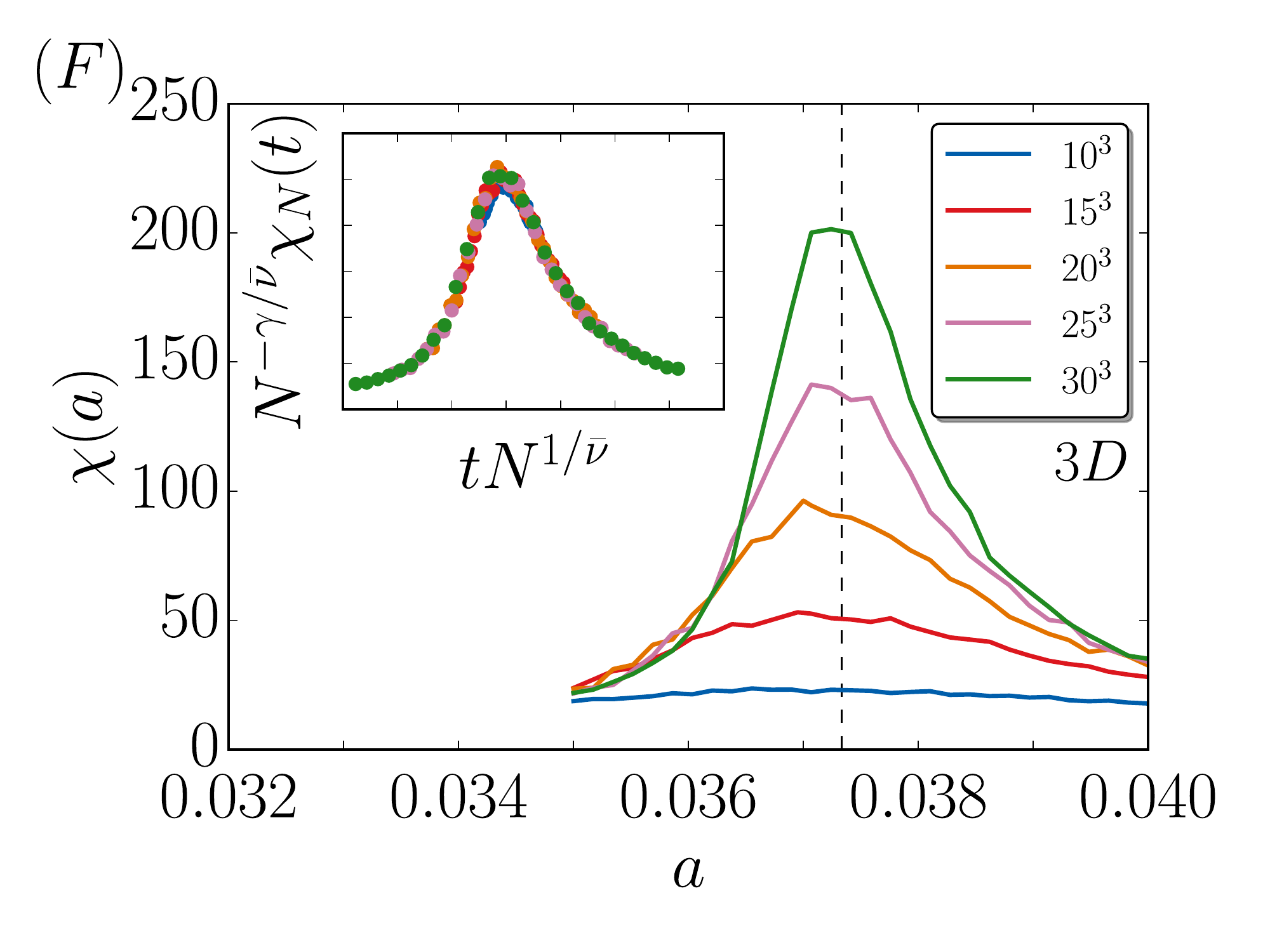}
\endminipage

\minipage{0.33\textwidth}
  \includegraphics[width=\linewidth]{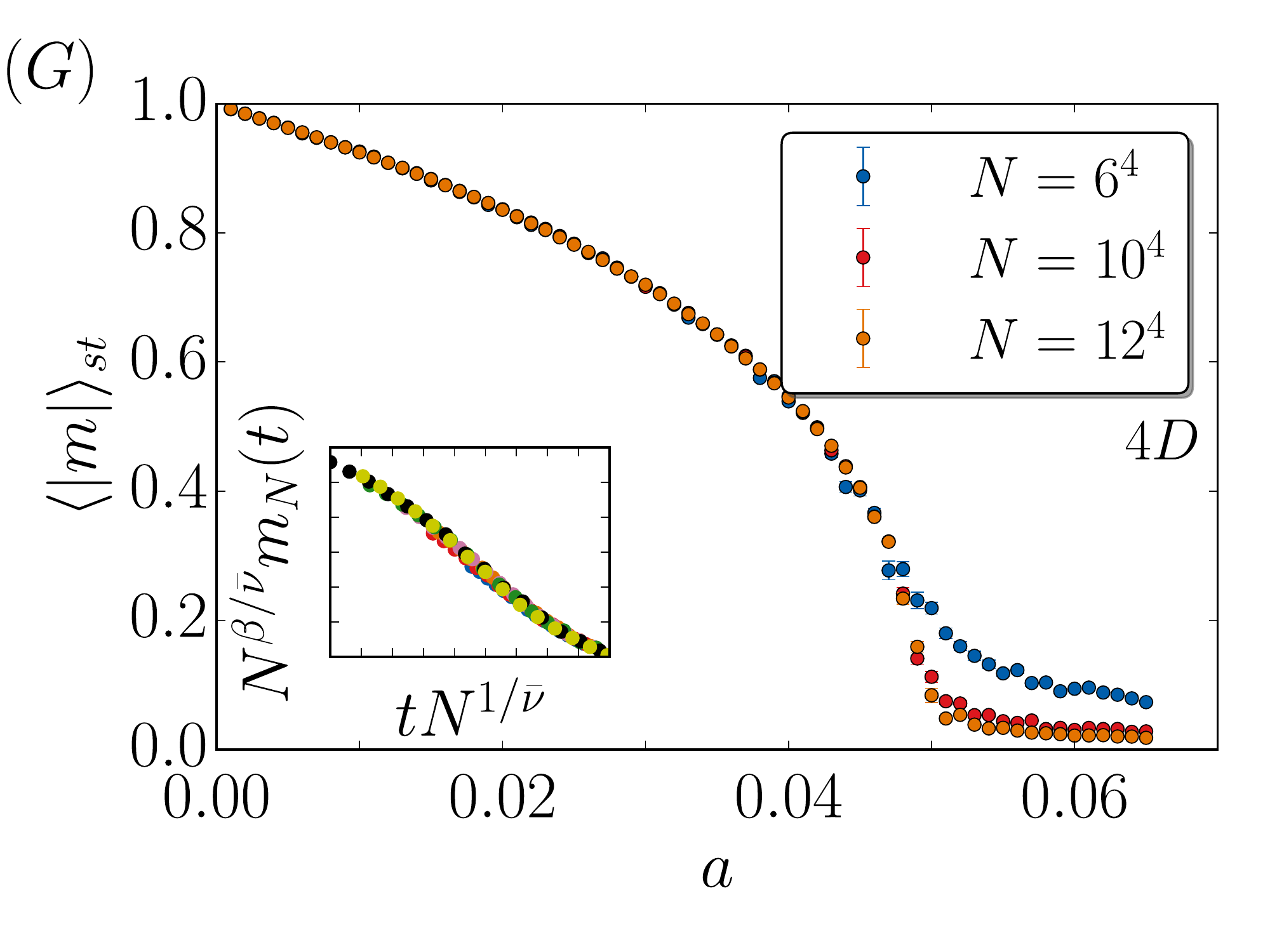}
\endminipage\hfill
\minipage{0.33\textwidth}
  \includegraphics[width=\linewidth]{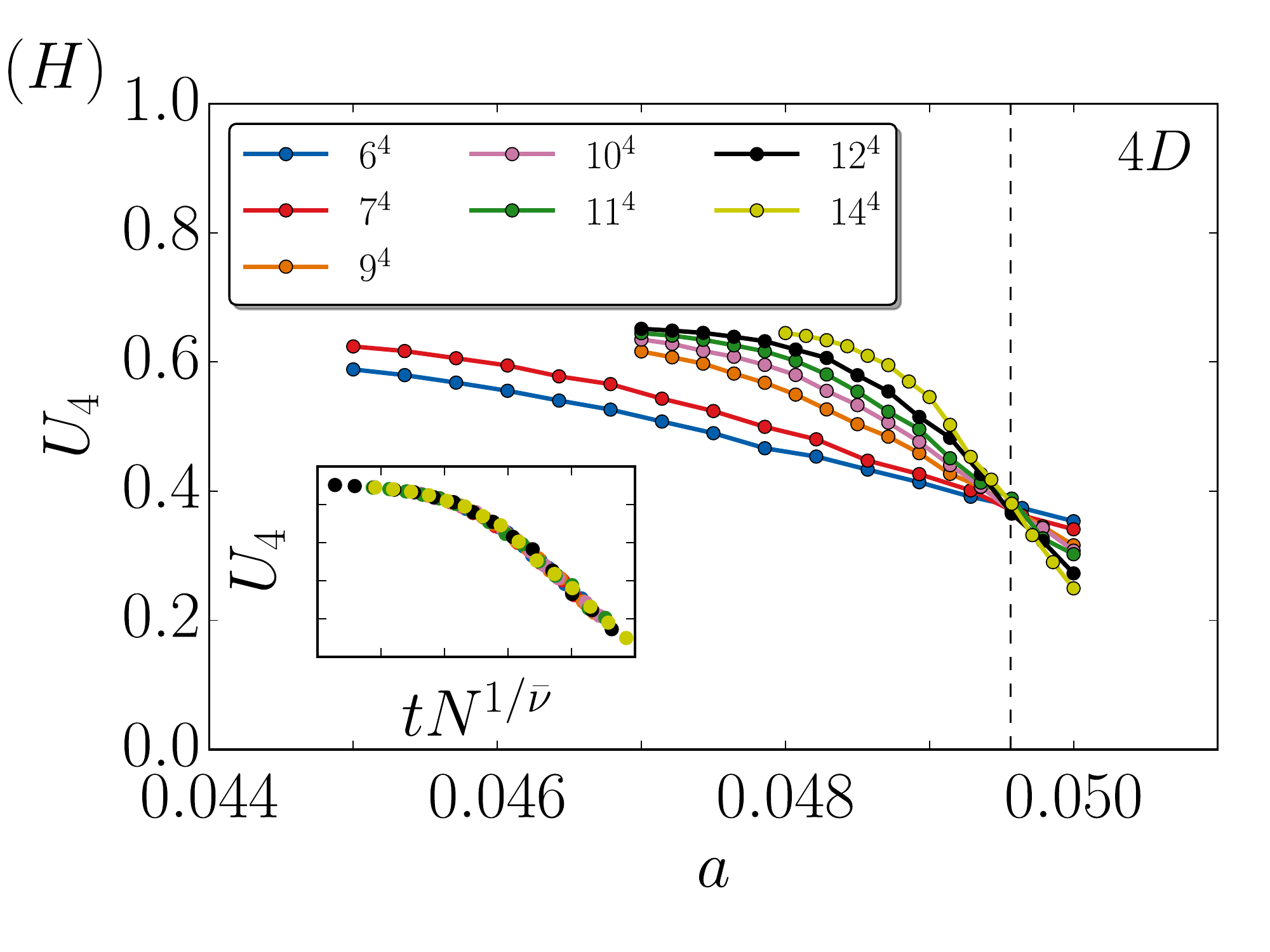}
\endminipage\hfill
\minipage{0.33\textwidth}%
  \includegraphics[width=\linewidth]{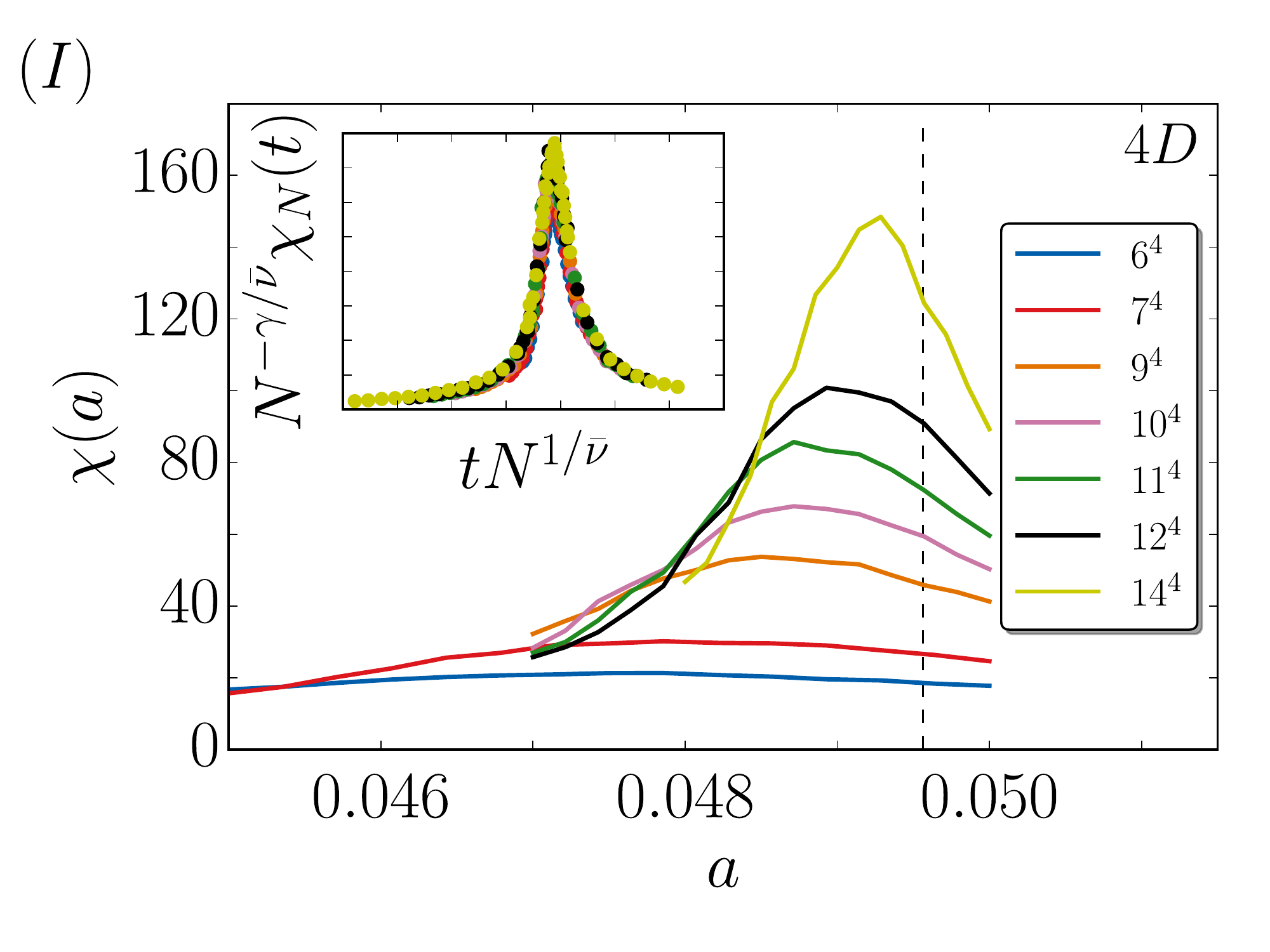}
\endminipage

\minipage{0.33\textwidth}
  \includegraphics[width=\linewidth]{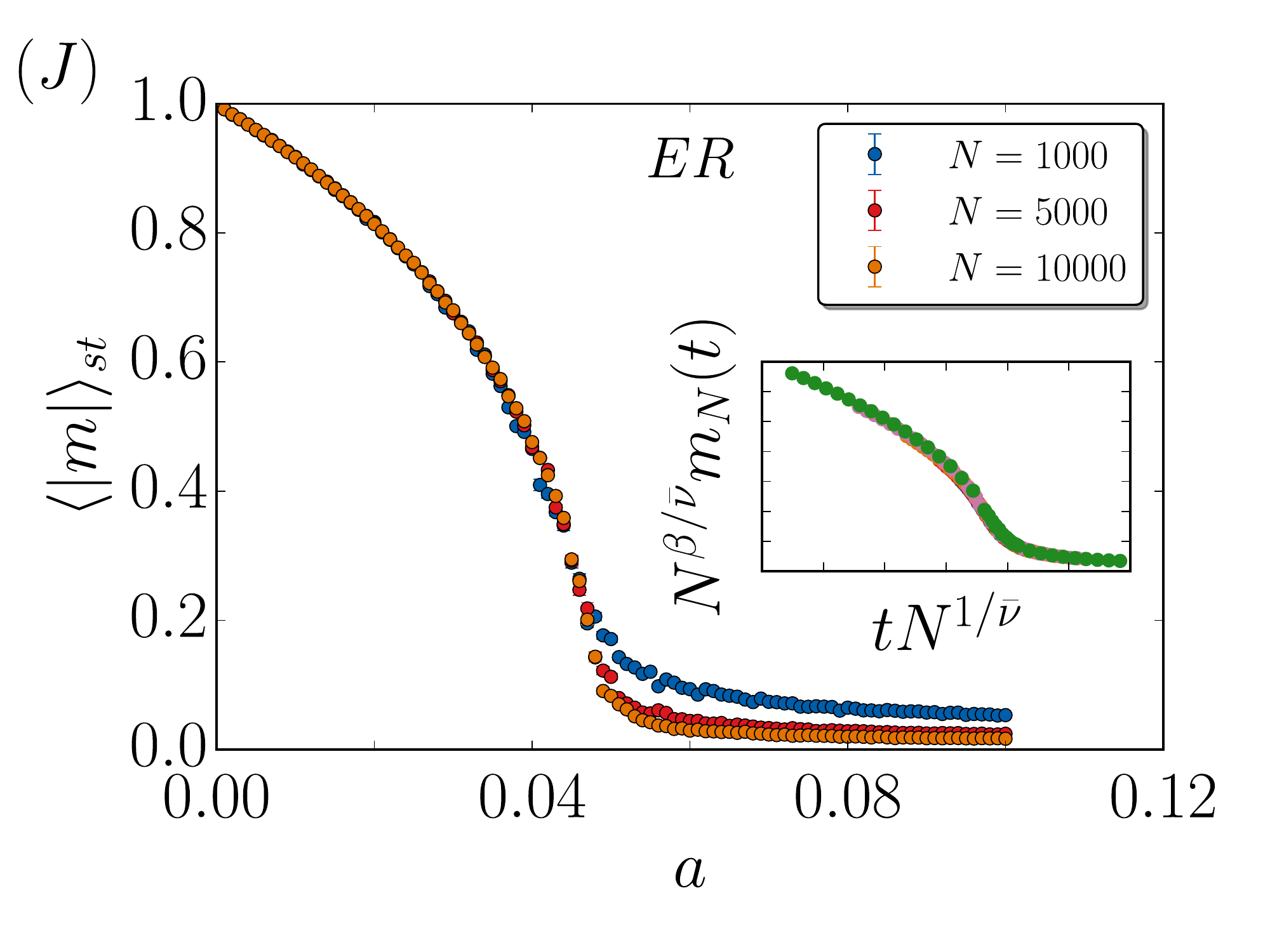}
\endminipage\hfill
\minipage{0.33\textwidth}
  \includegraphics[width=\linewidth]{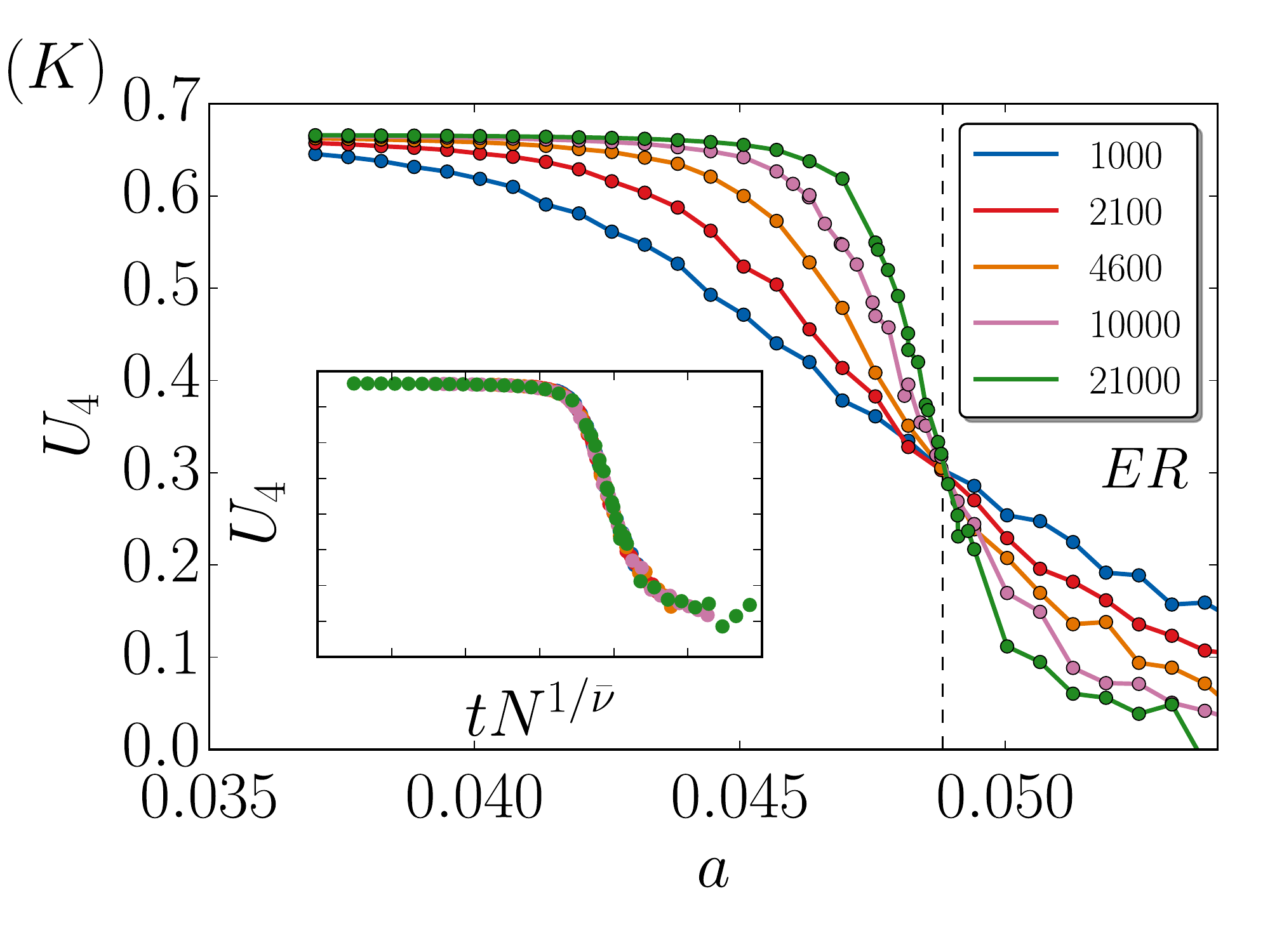}
\endminipage\hfill
\minipage{0.33\textwidth}%
  \includegraphics[width=\linewidth]{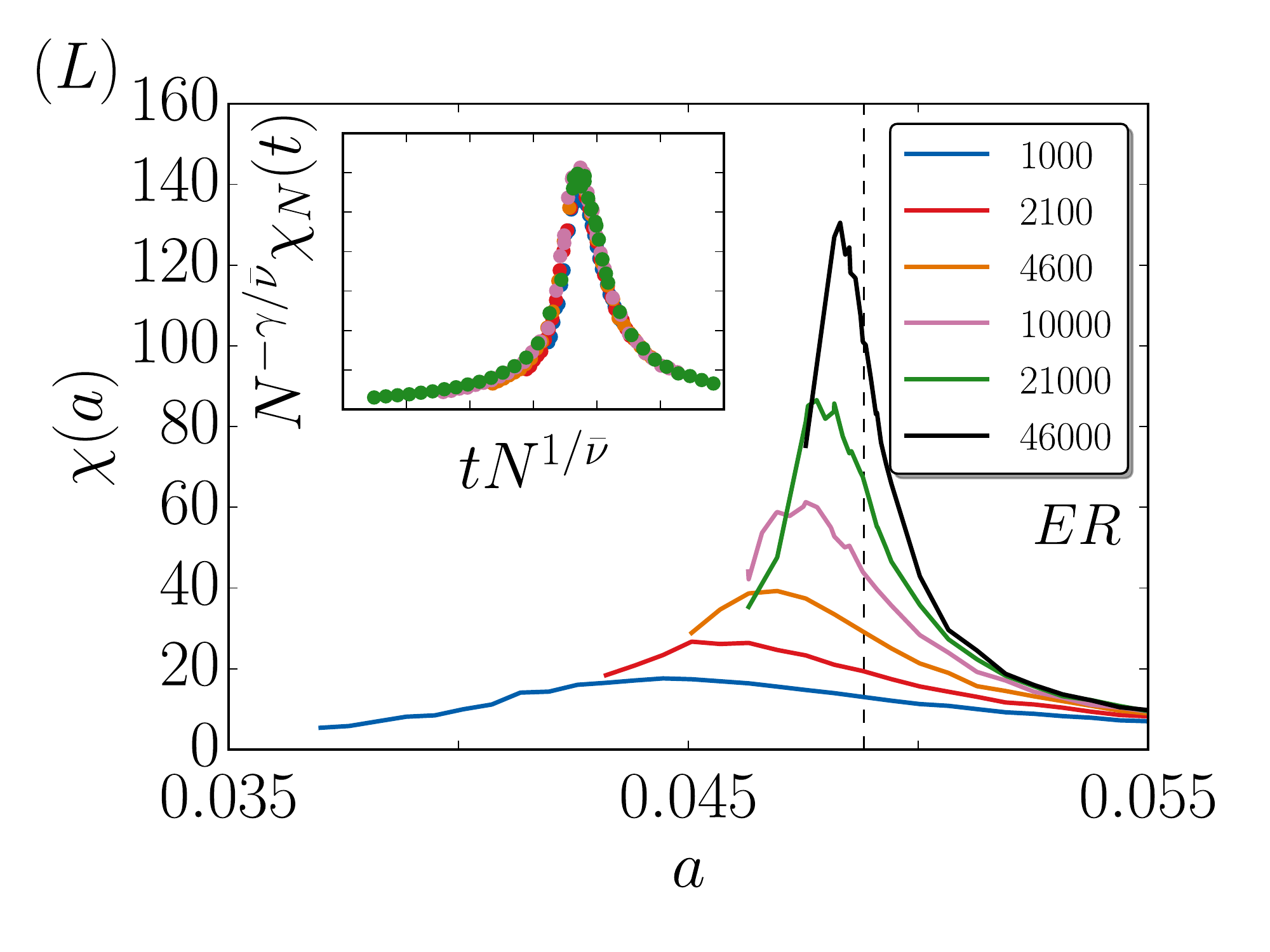}
\endminipage

\caption{Stationary magnetization ($A$, $D$, $G$, $J$), Binder cumulant ($B$, $E$, $H$, $K$), and susceptibility ($C$, $F$, $I$, $L$) for the noisy voter model with aging for different system sizes. From the top to the bottom row: lattices of $ d = 2 $, $ d = 3 $, $ d = 4 $ and Erd\H{o}s--R\'{e}nyi network with $ \langle k \rangle = 6 $. The insets show the collapses with the corresponding Ising critical exponents  \cite{yeomans1992statistical}. The vertical lines are located at the critical point to guide the eye.}\label{fig:Fig3}
\end{figure*}

\section{Noisy voter model with aging}
\label{sec:aging}

In the model with aging, each agent has her own internal time $ i = 0,1,2,...$ accounting for the time elapsed since her last change of state in Monte Carlo units. When it comes to pair--wise interactions, the node must be first activated with a probability $ 1/ (i+2) $. The motivation behind this choice is that the more an agent has spent in a given state (i.e., the larger $ i $), the more difficult is for her to change state via the voter update. The performance of the noisy update is not affected by aging. All nodes begin with their internal time equal to $ 0 $, so the effects of aging develop with the model evolution. When an agent changes state, due to either a noisy or a voter update, her internal time is reset to $ 0 $. On the contrary, this internal persistence time increases by one unit when the agent does not change state. This type of aging mechanism has been chosen because it induces features observed in several real-world systems, such as power-law interevent time distributions \cite{fernandez2011update}. 

Numerical simulations of the noisy voter model with aging are shown in Fig.~\ref{fig:Fig1}B. It is interesting to note that there is still a transition from bimodality to unimodality but of a completely different nature. The bimodality centered at $ \pm 1 $ occurs only for $ a = 0 $. As the noise increases, the peaks of $P_{st}(m)$ continuously move toward $m = 0$ [see Fig.~\ref{fig:Fig1}B]. The trajectories of the magnetization individually tend to a stationary $m$ value as shown in the inset of Fig.~\ref{fig:Fig1}B. The maxima of $P_{st}(m)$ continuously merge in a single peak for a given value of noise $a_c$. In contrast to the noisy voter model, the position of $a_c$ weakly depends on the system size [Fig.~\ref{fig:Fig1}C]. 

In the following we characterize the aging-induced continuous transition by finding a curve for the stationary magnetization. It is useful to introduce the variable $ n_i^+ $ (resp. $ n_i^- $), corresponding to the fraction of nodes in state $ 1 $ (resp. state $0$) and internal time $ i $ \cite{stark2008decelerating}. The total number of agents in state $ 1 $ is then $ n = \sum_{i=0}^{\infty} n_i^+ $, and equivalently $ N - n = \sum_{i=0}^{\infty} n_i^- $. We start by writing the transition rates that now depend on the internal time,
\begin{equation}
\begin{aligned}
\label{eq:3}
\Omega_1(i) & = n_i^+ \left( \frac{a}{2} + \frac{1-a}{2+i} \frac{N-n}{N} \right), \\
\Omega_2(i) & = n_i^- \left( \frac{a}{2} + \frac{1-a}{2+i} \frac{n}{N} \right), \\
\Omega_3(i) & = n_i^+ \left( \frac{a}{2} + \frac{(1-a)(1+i)}{2+i} + \frac{1-a}{2+i} \frac{n}{N} \right), \\
\Omega_4(i) & = n_i^- \left( \frac{a}{2} + \frac{(1-a)(1+i)}{2+i} + \frac{1-a}{2+i} \frac{N-n}{N} \right), 
\end{aligned}
\end{equation}
The first rate accounts for a node in state $ 1 $ changing state, thus reseting its internal time, i.e., $ n_i^+ \to n_i^+ - 1 $ and $n_0^- \to n_0^-+1$. The second rate is equivalent to the first but for a node in state $ 0 $. The third rate corresponds to the case of an agent in state $ 1 $ that is not able to change state, that is $ n_i^+ \to n_i^+ - 1 $ and $n_{i+1}^+ \to n_{i+1}^+ + 1$. It has several contributions: either a noisy update that does not result in a change in state, either the agent can not beat the aging when activating, or the agent overcomes the aging but it copies another node of her same state. The fourth rate is equivalent to the third but for a node in state $ 0 $. Note that $\Omega_1(i) + \Omega_3(i) = n_i^+$ and $\Omega_2(i) + \Omega_4(i) = n_i^-$.

We can write the temporal evolution of $\langle n_i^{\pm} \rangle$ as
\begin{align}
\frac{d \langle n_i^+ \rangle}{dt} & = - \langle \Omega_1(i) \rangle + \langle \Omega_3(i-1)\rangle - \langle \Omega_3(i)\rangle, \label{eq:req1} \\
\frac{d \langle n_i^- \rangle}{dt} & = - \langle \Omega_2(i)\rangle + \langle \Omega_4(i-1)\rangle - \langle \Omega_4(i)\rangle,   \label{eq:req2}
\end{align}
valid for times $ i \geq 1 $. For the particular case of $i = 0$,
\begin{align}
\frac{d \langle  n_0^+ \rangle}{dt} & = \sum_{i=0}^{\infty} \langle \Omega_2(i)\rangle - \langle\Omega_3(0)\rangle - \langle \Omega_1(0)\rangle, \label{eq:req3} \\
\frac{d \langle n_0^- \rangle}{dt} & = \sum_{i=0}^{\infty} \langle \Omega_1(i)\rangle - \langle \Omega_4(0)\rangle - \langle \Omega_2(0)\rangle,  \label{eq:req4}
\end{align}
with $ \langle \cdot \rangle $ standing for the average over realizations of the dynamics. Eqs.~\eqref{eq:req1}--\eqref{eq:req4} form an infinite set of coupled differential equations, which is hard to tackle analytically. However, valuable information can be extracted from the stationary solutions, obtained by setting the time derivative to $0$. 

By combining Eqs.~\eqref{eq:req1} and \eqref{eq:req3}, we have that
\begin{equation}
\label{eq:app2_1}
\frac{d \langle n \rangle }{dt} = 0 = \sum_{i=1}^{\infty} \langle \Omega_2(i) \rangle - \sum_{i=1}^{\infty} \langle \Omega_1(i) \rangle + \langle \Omega_2(0)\rangle - \langle \Omega_1(0)\rangle.
\end{equation}
Subtracting Eq.~\eqref{eq:req3} from Eq.~\eqref{eq:req4} and comparing with Eq.~\eqref{eq:app2_1}, we obtain the first condition for a stationary solution $ \langle \Omega_1(0)\rangle + \langle \Omega_3(0)\rangle = \langle \Omega_2(0)\rangle + \langle \Omega_4(0)\rangle $, which leads to $ \langle n_0^+ \rangle = \langle n_0^- \rangle $. That is, in the stationary regime the number of agents in states $ 1 $ and $ 0 $ that just reset their internal time is equal. The stationarity in Eq.~\eqref{eq:req1} leads to condition $ \langle \Omega_1(i) \rangle + \langle \Omega_3(i) \rangle = \langle \Omega_3(i-1) \rangle$. It is a recursive relation for $ \langle n_i^+ \rangle $, whose solution reads
\begin{align}
\label{eq:app2_2}
\langle n_i^+ \rangle & = \prod_{k=1}^{i} \left( \frac{a}{2} + (1-a)\frac{x+k}{1+k} \right) \langle  n_0^+ \rangle \nonumber \\
 & = \left(1-\frac{a}{2}\right)^i \frac{1}{\Gamma(2+i)} \left( \frac{2 + 2(1-a)x}{2-a}  \right)_i \langle  n_0^+ \rangle,
\end{align}
where $ x = \langle n \rangle/N $ and $ (z)_n \equiv z(z+1) \ldots (z+n-1)$ is the Pochhammer symbol. The stationarity in the second equation in \eqref{eq:req2} leads to the same equation for $ \langle n_i^- \rangle $ but with variables $ \langle  n_0^- \rangle $ and $ 1 - x $ instead of $ \langle n_0^+ \rangle $ and $ x $. Luckily, Eq. \eqref{eq:app2_2} and the corresponding one for nodes in state $0$ can be summed analytically so that we obtain the implicit equation,
\begin{align}
\label{eq:5}
\langle n \rangle & = \frac{1}{a+2(1-a)x} \left( 2^{\frac{-2+2(a-1)x}{a-2}} a^{\frac{a+2(1-a)x}{a-2}} - 2 \right) \langle  n_0^+ \rangle \nonumber \\
& \equiv f(a,x) \langle  n_0^+ \rangle,
\end{align} 
where $ x = \langle n \rangle/N $ and the function $ f(a,x) $ has been introduced to ease the notation. Another condition of stationarity leads to the very same equation but with variables $ \langle  n_0^- \rangle $ and $ 1 - x $ instead of $ \langle n_0^+ \rangle $ and $ x $, i.e., $ N - \langle n \rangle = f(a,1-x) \langle n_0^- \rangle$. Using the condition that $ \langle  n_0^+ \rangle = \langle  n_0^- \rangle $, we find that 
\begin{equation}
\label{eq:6}
\frac{x}{1-x} = \frac{f(a,x)}{f(a,1-x)},
\end{equation} 
whose solutions $ x(a) $ give the noise dependent curves of the magnetization in the stationary state, also called the equation of state. One trivial solution is $ x = 1/2 $, which can be checked by direct substitution in Eq. \eqref{eq:6}. This corresponds to the symmetric case with equal coexistence of agents in both states. It is a stable solution for $ a > a_c $ and unstable for $a < a_c$. Moreover, for $ a < a_c $, two new stable and symmetrical solutions appear, corresponding to the two ferromagnetic branches.

At the critical noise $ a_c $, the derivatives with respect to $ x $ on the two sides of Eq. \eqref{eq:6} coincide when evaluated at $ x = 1/2 $. After simple but lengthy algebra, one obtains the equation for the critical point,
\begin{equation}
\label{eq:7}
\frac{(2-a_c)^2}{1-a_c} = \log \left(\frac{2}{a_c} \right) \left( 1 - \left( \frac{a_c}{2}\right)^{\frac{1}{2-a_c}} \right)^{-1},
\end{equation}
which gives $ a_c = 0.07556 ... $ With this information, we can readily obtain the critical exponent of the magnetization $ \beta $ such that $ m \sim t^{\beta} $, where $ t \equiv (1 - a/a_c) $ is the reduced noise.

Since $ m = 2x -1 $, the behavior of $ x $ close to the critical point will be the same as $ m $. We Taylor expand the two sides of Eq. \eqref{eq:6} around $ x = 1/2 $ and $ a = a_c $, so
\begin{align}
& \frac{x}{1-x} \sim 1 + 4 (x - 1/2) + 8 (x - 1/2)^2 + 16 (x - 1/2)^3 \nonumber \\
& \hskip70pt + \mathcal{O}(x^4) \nonumber \\
& \frac{f(a,x)}{f(a,1-x)} \sim 1 + (4 - 21.1 (a - a_c))(x - 1/2) \nonumber \\
& \hskip70pt + 8 (x - 1/2)^2 + 10.5 (x - 1/2)^3 \nonumber \\
& \hskip70pt + \mathcal{O}(x^4,a^2)
\end{align}
Equating both sides, the smallest nonvanishing order is $ \beta = 1/2 $.

So far the results are in the form of mean field analytical expressions, and we computationally verify next their validity and study the finite size scaling. The numerical simulations for the equation of state of $|m|$ is displayed in Fig.~\ref{fig:Fig2}A for different system sizes along with the theoretical solution $ x(a) $. To test the accuracy of the prediction for the critical point $ a_c $, we can employ the technique of the Binder cumulant computing  $ U_4(a) = 1 - \langle m^4 \rangle_{st} / 3 \langle m^2 \rangle_{st} ^2  $ \cite{binder1981finite}, being $\langle m^{n} \rangle_{st} $ the $ n $th moment of the magnetization in the stationary state. This provides an accurate estimate of the critical noise as the crossing point of the Binder cumulant curves for different $N$'s [Fig.~\ref{fig:Fig2}B]. We obtain $ a_c = 0.0753(6)$, which is compatible with the mean field theoretical value. It is important to note that we can define neither an internal energy nor a specific heat since the aging noisy voter model does not possess a Hamiltonian. Nevertheless, the susceptibility $ \chi(a) = N (\langle m ^2 \rangle_{st} - \langle m \rangle_{st} ^2 )$, understood as the fluctuations in the magnetization, is well defined. In Fig.~\ref{fig:Fig2}C, we plot the susceptibility for different system sizes and confirm that $\chi(a) $ diverges at $ a_c $ when $ N \to \infty $.

We use the techniques of finite-size scaling to collapse the data and determine the critical exponents of this system. The scaling hypothesis predicts that close to the critical point (i.e., $ t = 1 - a/a_c \to 0 $) the magnetization behaves as $m = N^{-v} f_{1} (t\, N^u)$ and the susceptibility as $\chi = N^w \, f_{2} (t\, N^u)$ with $ v = \beta/d \, \nu $, $ u = 1/ d \, \nu $ and $ w = \gamma/d \, \nu $, where $ d $ is the dimension in which the system is embedded and $ \gamma $ and $ \nu $ are the critical exponents for the susceptibility and the correlation length. Above the critical dimension $d_c$, the exponents become mean field, and they do not depend on $d$ anymore. Compactly, we write that $ v = \beta/\bar{\nu} $, $ u = 1/ \bar{\nu} $ and $ w = \gamma/\bar{\nu} $ where $\bar{\nu} = d_c \, \nu $ in the case that $ d \geq d_c $, otherwise $\bar{\nu} = d \, \nu $ \cite{deutsch1992optimized}. This has a direct consequence when studying phase transitions of a system above its critical dimension---as in an all-to-all topology, such as the present case---we can have access to $ \beta $ and $ \gamma $ but not to $ \nu $. What one obtains is the quantity $ \bar{\nu} $, so in order to know the actual value of $ \nu $, we must know the critical dimension $ d_c $ beforehand. 

\begin{figure}
  \includegraphics[width=0.8\linewidth]{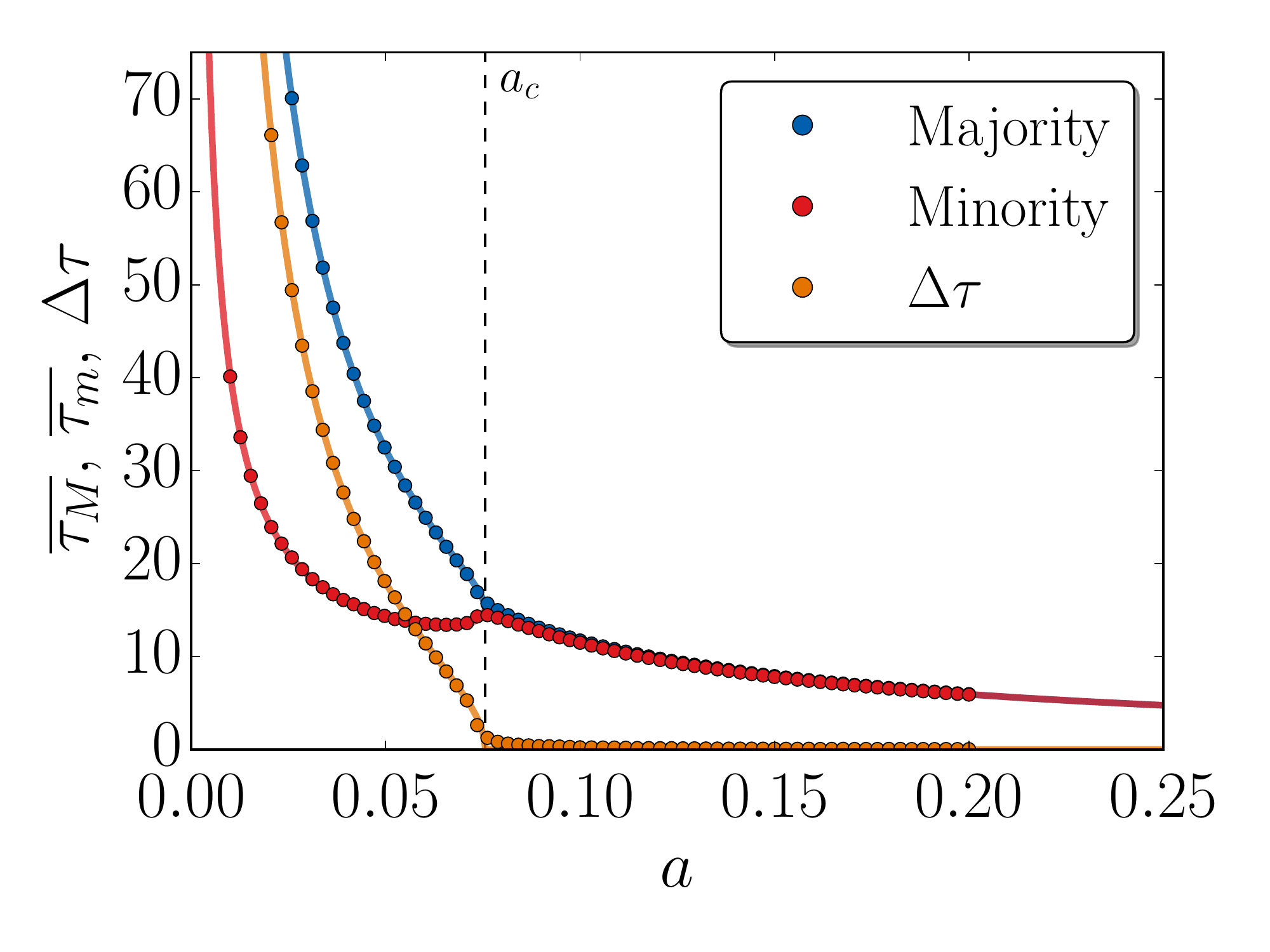}
\caption{Mean internal times and their difference $ \Delta \tau $. The points are obtained from simulations, and the solid lines are obtained from theory \cite{footnote}.}\label{fig:Fig4}
\end{figure}
We analytically proved that $ \beta = 1/2 $ for the noisy voter model with aging in complete graphs. It coincides with the mean field exponent of the Ising model, so this universality class is a reasonable candidate for our model. In the mean field regime of this class $ \gamma = 1 $, $ \bar{\nu} = 2 $ and the critical dimension is $ d_c = 4 $. By data collapsing the magnetization, the Binder cumulant, and the susceptibility [insets of Fig.~\ref{fig:Fig2}A--\ref{fig:Fig2}C] we confirm that $\beta$,  $ \gamma$, and $ \bar{\nu}$ take these values, although we cannot establish $d_c$ from an all-to-all framework as discussed. To proceed, we can compute the same quantities in lower dimensions and collapse the curves using the corresponding Ising critical exponents assuming $ d_c = 4 $. In case the collapses neatly overlap, we can conclude that the Ising universality class is a solid candidate for the noisy voter model with aging. We show these analyses for lattices of dimensions $ d = 2, \, 3, $ and $ 4 $ and Erd\H{o}s--R\'{e}nyi networks, which have an effective infinite dimensionality, in Fig. \ref{fig:Fig3}. We find excellent overlaps, ratifying, thus, that our system is compatible with the universality class of the Ising model.

An explanation of the mechanism behind aging-induced phase transition can be given in terms of symmetry breaking associated with the internal times of the nodes. In the disordered (paramagnetic) phase there is no predominant state since the dynamics is driven mainly by noise: In this case all nodes are of similar age. In the ferromagnetic phase, the system is ordered towards one of the states, displaying a net magnetization. In this scenario, noisy updates are less frequent than voter ones, which take into account the neighbor states. If there is a global majority opinion, then nodes holding this opinion will flip less times than those of the minority. This introduces an asymmetry in the age distribution of the nodes. Consequently, on average we have that older nodes belong to the majority in the region $ a < a_c $. These arguments can be quantified by using the average age of each node population. The mean internal time \cite{artime2017joint} of the majority population is $\overline{\tau}_{M} = \sum_{i} i \langle n_i^{M} \rangle / \sum_{i} \langle n_i^{M} \rangle $ where the index $ M $ is $ + $ or $-$ depending on which population dominates. A complementary expression follows for the mean internal time of the minority population $\overline{\tau}_{m}$. Fig.~\ref{fig:Fig4} shows these quantities: In the region $ a < a_c $ the asymmetric aging in the populations is evident by the separation in the upper branch (older nodes, belonging to the majority) and the lower branch (younger nodes, in the minority). These two branches merge at the critical point $ a = a_c $. In the disordered phase $ a > a_c $, there are similar numbers of nodes in state $ 1 $ and in state $ 0 $ and the internal times of both populations are the same. Thus, $\Delta \tau = |\overline{\tau}_{M} - \overline{\tau}_{m}| $ can be used as an alternative order parameter: In the paramagnetic phase $\Delta \tau = 0$, and in the ferromagnetic one $\Delta \tau \neq 0$.

\section{Discussion}

To summarize, we have explored the effect of aging in a stochastic binary-state model. The agents have now an internal time counting the time spent in the same state, with older agents less prone to update. When aging is added to the noisy voter (Kirman) model, we find numerically and analytically that the system passes from a discontinuous transition, whose transition point vanishes in the thermodynamic limit, to a robust second order phase transition of the Ising universality class.
Indeed, the Ising model gives some intuition to understand the reported phenomena in this article: It is a model that at temperature $ T = 0 $ coarsens and its ergodicity is broken due to its absorbing states. When the temperature is finite, the ergodicity is restored, and, at a critical temperature $ T = T_c $, the system displays a continuous phase transition. Similarly, incorporating aging into the voter model generates an algebraic coarsening dynamics and an ergodicity breaking as shown in Ref. \cite{fernandez2011update} for complete and random graphs. Hence, the combination of aging and noise (with a similar role as the temperature in the Ising model) induces a continuous phase transition at a well-defined critical point, despite the coarsening characteristics in both models can be different. Therefore, one can hypothesize that adding a mechanism that produces coarsening, such as aging, in a model in which random spin flips are allowed may lead to an Ising-like phase transition. This conjecture, observed here, requires further research to be generalized. Finally, we show how aging properties of the agents can be employed to understand the spontaneous symmetry breaking between states below the critical point, proving that aging plays a central role in modifying the critical properties of a system.

\section*{Acknowledgments}

Partial financial support has been received from the Agencia Estatal de Investigacion (AEI, Spain) and Fondo Europeo de Desarrollo Regional under Project ESOTECOS Project No. FIS2015-63628-C2-2-R (AEI/FEDER,UE) and the Spanish State Research Agency, through the Maria de Maeztu Program for units of Excellence in R\&D (MDM-2017-0711). A.F.P. acknowledges support by the Formacion de Profesorado Universitario (FPU14/00554) program of Ministerio de Educacion, Cultura y Deportes (MECD) (Spain). We thank Juan Fern\'andez-Gracia for a careful reading of the paper and useful comments.

\bibliography{bibliografia.bib}

\begin{thebibliography}{49}
\bibitem [{foo()}]{footnote}%
  \BibitemOpen
  \href@noop {} {\ }\bibinfo {note} {The mean internal time of nodes in state $
  1 $ can be computed as $\overline{\tau}_{+} = \frac{\sum_{i} i \langle
  n_{i}^{+}\rangle}{\sum_{i} \langle n_{i}^{+}\rangle} = 2
  \frac{2^{\frac{2(1+x)}{a-2}} a^{\frac{2 a x}{a - 2}}+2^{\frac{2 a x}{a-2}}
  a^{\frac{2(1+x)}{a-2}} x - 2^{\frac{2 a x}{a-2}} a^{\frac{a+2x}{a-2}}
  x}{2^{\frac{2 a x}{a-2}} a^{\frac{a+2x}{a-2}} - 2^{\frac{a+2x}{a-2}}
  a^{\frac{2 a x}{a-2}}}$. The expression for nodes in state $ 0 $ can be
  easily obtained after substituting $x$ by $1-x$ in the last
  equation.}\BibitemShut {Stop}%
\end{thebibliography}

\end{document}